\documentclass[journal,twoside,web]{ieeecolor}
\usepackage{tmi}
\usepackage{cite}
\usepackage{amsmath,amssymb,amsfonts}
\usepackage{algorithmic}
\usepackage{graphicx}
\usepackage{textcomp}
\usepackage{times}
\usepackage{epsfig}
\usepackage{subfigure}
\usepackage{verbatim}
\usepackage{ragged2e}
\usepackage{booktabs}
\usepackage{multicol}
\usepackage{gensymb}
\usepackage{hyperref}
\usepackage{tabularx}
\usepackage{overpic}
\usepackage[switch]{lineno}
\usepackage{threeparttable}
\usepackage{multirow}
\usepackage{mathrsfs,bbm}
\usepackage{float}

\usepackage[draft]{todonotes}   % notes shown

\def \ourmodel{C$^{2}$DA-Net}

\def\BibTeX{{\rm B\kern-.05em{\sc i\kern-.025em b}\kern-.08em
    T\kern-.1667em\lower.7ex\hbox{E}\kern-.125emX}}
\begin{document}
%\linenumbers

% \title{Unsupervised Domain Adaptation for Clinical Fetal Brain Tissue Segmentation via Cycle-Consistent Networks}
\title{Domain Adaptive Segmentation of Clinical Fetal Brain MR Images via Cycle-Consistent Networks} % Other potential titles.
% \title{Segmentation of Clinical Fetal Brain MR Images via a Domain Adaptation Inspired Cycle-Consistent Networks} % Other potential titles.
\title{Unsupervised Tissue Segmentation of Clinical Fetal Brain MR Images via a Fourier Domain Adapt-Cycle Network} % Other potential titles.
% \title{Unsupervised Segmentation of Style-Removed Clinical Fetal Brain MR Images via Domain Adaptive Cycle-Consistent Networks}
% \title{Learning Segmentation from Style-Removed isotropic reconstructed high-quality Images via Domain Adaptive Cycle-Consistent Networks}
% \title{Frequency Decomposition and Cycle Consistency Guided Domain Adaptive Network for Clinical Fetal Brain Tissue Segmentation}
% \title{Frequency Decomposition Guided Domain Cycle-Adapt Network for Clinical Fetal Brain Tissue Segmentation}
% \title{Domain Cycle-Adapt Network for Clinical Fetal Brain Tissue MRI Segmentation}
% \title{High-Quality Guided Clinical Fetal Brain Tissue Segmentation via a Domain Adaptation Network Inspired Cycle-Consistent} % Other potential titles.
\title{High-Quality Guided Clinical Fetal Brain Tissue Segmentation via a Cycle Domain Adaptation Network} % Other potential titles.
% \title{Learning Tissue Segmentation from High-Quality Reconstructed Fetal Brain Images} % Other potential titles.
\title{High-Quality Guided Clinical Fetal Brain Tissue MRI Segmentation} % Other 
\title{Learning thick-slice Scans Segmentation From Isotropic Reconstructed Fetal Brain MR Images}
\title{Learning Tissue Segmentation From Isotropic Reconstructed Fetal Brain MR Images for Thick-Slice Scans}
\title{Learning Tissue Segmentation of Fetal Brain Thick-Slice Scans From Isotropic Reconstructed MR Images}
\title{Tissue Segmentation of Thick-Slice Fetal Brain MR Scans with Guidance from High-Quality Isotropic Volumes}

\author{Shijie Huang,
       Xukun Zhang,
       Zhiming Cui,
       He Zhang,
       Geng Chen,
       Dinggang Shen \IEEEmembership{Fellow, IEEE} 
        % and~Jane~Doe,~\IEEEmembership{Life~Fellow,~IEEE}% <-this % stops a space
% \thanks{IDEA Lab}% <-this % stops a space
% \thanks{J. Doe and J. Doe are with Anonymous University.}% <-this % stops a space
% \thanks{Manuscript received April 19, 2005; revised August 26, 2015.}
\thanks{Shijie Huang and Zhiming Cui are with the School of Biomedical Engineering, ShanghaiTech University, Shanghai, China (e-mail: huangshj@shanghaitech.edu.cn; zmcui@cs.hku.hk).}
\thanks{Xukun Zhang is with Academy for Eng. \& Tech. Fudan University, Shanghai, China (e-mail: zhangxk21@m.fudan.edu.cn).}
\thanks{He Zhang is with the Department of Radiology, Obstetrics and Gynecology Hospital of Fudan University, Shanghai, China (dr.zhanghe@yahoo.com)}
\thanks{Geng Chen is with the National Engineering Laboratory for Integrated Aero-Space-Ground-Ocean Big Data Application Technology, School of Computer Science and Engineering, Northwestern Polytechnical University, Xi'an, China (geng.chen.cs@gmail.com)}
\thanks{Dinggang Shen is with the School of Biomedical Engineering, ShanghaiTech University, Shanghai 201210, Shanghai United Imaging Intelligence company Ltd., Shanghai 200030 and also Shanghai Clinical Research and Trial Center, Shanghai 201210, China (e-mail: Dinggang.Shen@gmail.com).}
}
\maketitle

\begin{abstract}
Accurate tissue segmentation of thick-slice fetal brain magnetic resonance (MR) scans is crucial for both reconstruction of isotropic brain MR volumes and the quantification of fetal brain development.
However, this task is challenging due to the use of thick-slice scans in clinically-acquired fetal brain data.
To address this issue, we propose to leverage high-quality isotropic fetal brain MR volumes (and also their corresponding annotations) as guidance for segmentation of thick-slice scans.
Due to existence of significant domain gap between high-quality isotropic volume (i.e., source data) and thick-slice scans (i.e., target data), we employ a domain adaptation technique to achieve the associated knowledge transfer (from high-quality $<$source$>$ volumes to thick-slice $<$target$>$ scans).
Specifically, we first register the available high-quality isotropic fetal brain MR volumes across different gestational weeks to construct longitudinally-complete source data.
To capture domain-invariant information, we then perform Fourier decomposition to extract image content and style codes.
Finally, we propose a novel Cycle-Consistent Domain Adaptation Network (\ourmodel) to efficiently transfer the knowledge learned from high-quality isotropic volumes for accurate tissue segmentation of thick-slice scans.
Our \ourmodel{} can fully utilize a small set of annotated isotropic volumes to guide tissue segmentation on unannotated thick-slice scans.
Extensive experiments on a large-scale dataset of 372 clinically acquired thick-slice MR scans demonstrate that our \ourmodel{} achieves much better performance than cutting-edge methods quantitatively and qualitatively.

\end{abstract}

% Note that keywords are not normally used for peerreview papers.
\IEEEpeerreviewmaketitle
\section{Introduction}
\begin{figure}[t]
  \centering
  \includegraphics[width=\linewidth]{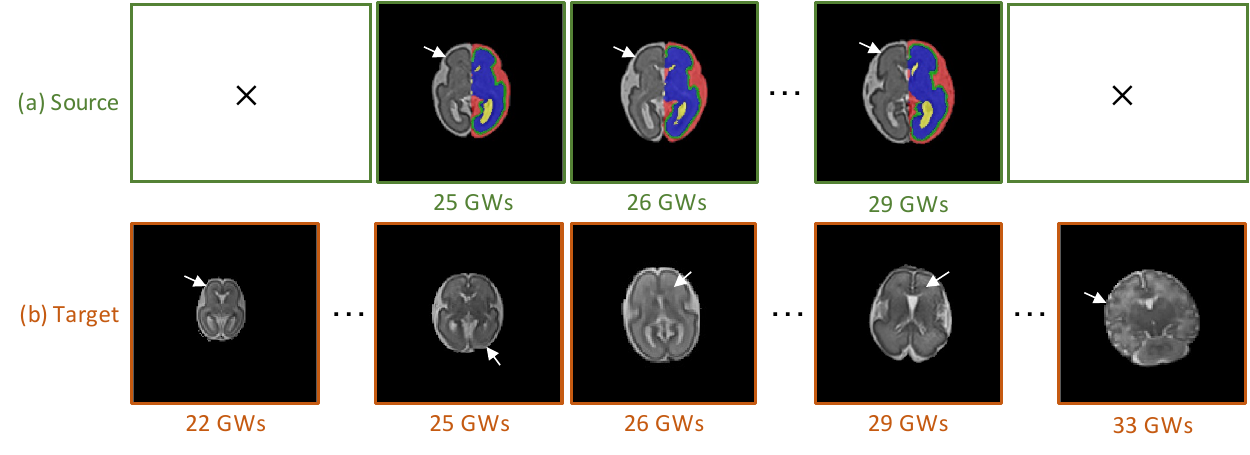}
  % \vspace{-20pt}
  \caption{Typical examples of target and source data. (a) Reconstructed high-quality isotropic MR volumes (voxel size: 0.75$\times$0.75$\times$0.75 (mm)) with six kinds of annotations for different tissues, including CSF (Cerebrospinal Fluid), GM (Gray Matter), WM (White Matter), ventricles, cerebellum, and brainstem. (b) Clinical-acquired thick-slice MR scans (voxel size: 0.75$\times$0.75$\times$4.4 (mm)) with low quality. Reconstructed images are missing some GWs}
  \label{fig0}
\end{figure}

% 1. What is clinical tick-slice MR scan? 2. Importance of tissue segmentation of tick-slice MR scan; 3. Why it is challenging?
\IEEEPARstart{N}{on-invasive} fetal magnetic resonance (MR) imaging is an essential technique for early fetal examination, e.g., fetal brain development \cite{huang2021handbook,jokhi2011magnetic,gholipour2017normative,kuklisova2012reconstruction}.
Clinically, fetal MR scans are acquired as thick-slice stacks, and then reconstructed to generate high-resolution isotropic brain MR volumes.
Segmentation of fetal brain tissues from these thick-slice MR scans is critical for numerous downstream tasks, such as super-resolution reconstruction of isotropic MR volumes, which can benefit from the anatomical information provided by the segmentation.
However, even with advanced deep learning techniques, manual annotation and segmentation of fetal brain tissue from thick-slice MR scans remain challenging \cite{jokhi2011magnetic,zhang2021confidence} due to the blurry boundary between the cortical plate and the cerebrospinal fluid, as shown in Fig. \ref{fig0} (b), as well as the anatomic specificity of the developing cerebrum during different gestational weeks (GWs) \cite{huang2021handbook,gholipour2017normative}. As a result, manual annotations may be error-prone and unreliable for training deep neural networks, leading to poor segmentation and analysis performance.

Significant efforts have been dedicated to fetal brain tissue segmentation. Existing methods primarily focus on the segmentation of high-quality isotropic MR volumes.
For instance, Gholipour \emph{et al.} \cite{gholipour2017normative} and Dumast \emph{et al.} \cite{de2022synthetic} proposed traditional and deep learning methods for fetal brain tissue segmentation of isotropic MR volumes.
Furthermore, challenges \cite{payette2022fetal} based on isotropic MR volumes have been held to compare fetal brain tissue segmentation methods.
However, a major limitation of these methods is that they require high-quality isotropic reconstructed MR volumes, which are not always clinically available.
Furthermore, these methods ignore the significant potential of thick-slice fetal brain tissue segmentation to facilitate downstream tasks such as reconstruction, registration, quantitative analysis, etc.
Therefore, there is a great need to develop an automatic method for accurately segmenting fetal brain tissues from thick-slice scans. Compared with clinical-acquired scans, the reconstructed fetal brain MR volumes have higher quality, isotropic resolution, and more accurate segmentation annotations, providing key information to guide the segmentation of thick-slice fetal brain MR scans.

Inspired by this observation, we propose to segment thick-slice fetal brain MR scans with guidance from high-quality isotropic volumes.
To address the large gap between high-quality isotropic volume (i.e., source data) and thick-slice scans (i.e., target data), we employ a domain adaptation technique to achieve the associated knowledge transfer (from high-quality $<$source$>$ volumes to thick-slice $<$target$>$ scans).
In addition to this domain gap, we describe fetal development as a special domain shift, which is alleviated by building longitudinal-complete source data.
Specifically, we first register high-quality isotropic volumes across GWs to complement the longitudinal missing ones.
The Fourier transformation is then introduced to decompose all images according to their frequency maps, providing Fourier Content Code (FCC) and Fourier Style Code (FSC) that represent high-frequency domain-invariant content and low-frequency domain-specific style, respectively.
Finally, we propose a Cycle-Consistent Domain Adaptation Network (\ourmodel) to learn domain-invariant structure from FCC and FSC for tissue segmentation.
By leveraging a small set of annotated isotropic reconstructed MR volumes to guide brain tissue segmentation on the unannotated thick-slice MR scans, our method overcomes the challenges of the large domain gap and the special domain shift.
The main contributions of our work are summarized as follows:
\begin{itemize}
    \item We propose to leverage prior knowledge from reconstructed high-resolution high-quality MR volumes to guide the brain tissue segmentation of clinically-acquired thick-slice MR scans.
    \item To bridge the significant gap between the two domains, we employ the domain adaptation technique to achieve the associated knowledge transfer, which is effectively resolved with a registration-based longitudinal source data completion and a specially-designed cycle-consistent network, i.e., \ourmodel{}.
    \item Our \ourmodel{} (i) incorporates Fourier decomposition to capture both domain-invariant and domain-specific components, and (ii) introduces a cycle-consistent constraint to ensure anatomical accuracy of features.
    \item We extensively evaluate \ourmodel{} on a large-scale clinical-acquired fetal MRI dataset, and our results demonstrate that it achieves promising performance and outperforms state-of-the-art methods significantly.
\end{itemize}

\begin{figure*}[t]
  \centering
  \includegraphics[width=\linewidth]{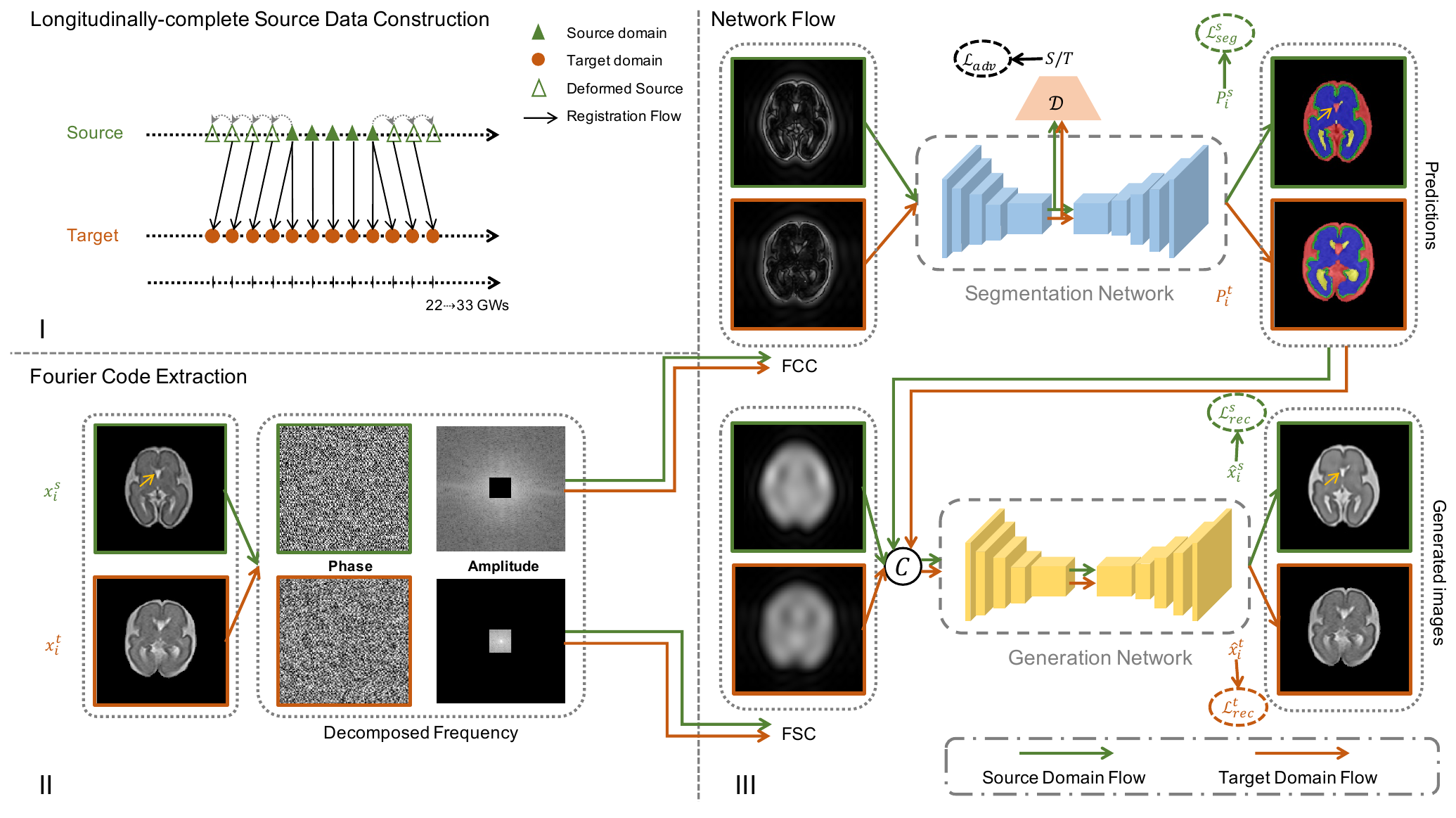}
  \vspace{-16pt}
  \caption{Overview of the proposed method for thick-slice fetal brain tissue segmentation.
  (I) We first register the source images to the target images at similar GWs to build deformed longitudinal-complete source data. (II) We then extract the FCC and FSC from the input images. (III) The network flow includes a segmentation network ($S$) and a generation network ($G$). An FSC cycle-consistence network trains the MR images in an unsupervised manner, where feature space alignments are employed.
  }
  \label{fig2}
\end{figure*}

\section{Related Work}
\subsection{Fetal Brain Tissue Segmentation}
Tissue segmentation of fetal MR images is crucial to investigate structural changes in brain development, detect brain injuries and malformations, and provide prognostic information \cite{huang2021handbook}.
Several brain tissue segmentation methods have been proposed for fetal subjects with isotropic reconstructed MR volumes.
Dumast \emph{et al.} \cite{de2022synthetic} synthesized annotated data that can be used for domain adaptation, significantly boosting the segmentation performance of seven brain tissues.
Following a similar idea, Li \emph{et al.} \cite{li2022cas} jointly generated conditional atlases for brain segmentation.
In particular, Payette \emph{et al.} \cite{payette2022fetal} organized the Fetal Tissue Annotation (FeTA) Challenge, where 20 international teams participated and submitted a total of 21 algorithms for evaluation, significantly boosting the development of fetal brain tissue segmentation for isotropic reconstructed high-quality MR volumes.

Despite the progress in this field, most methods are designed for isotropic reconstructed high-quality fetal brain MR volumes, which are not always available in a clinical setting.
Until now, little attention has been paid to segmenting brain tissues directly from thick-slice fetal scans that can assist downstream tasks such as registration, reconstruction, and neurodevelopment analysis \cite{gholipour2017normative,kuklisova2012reconstruction}.

\subsection{Unsupervised Domain Adaptation}
Domain adaptation is a widely-used approach in transfer learning that aims to improve performance when there is a domain gap between the source and target data.
In medical image analysis, the domain gap is mainly caused by cross-modality or inter-scanner variations \cite{cui2021structure}.
Early research focused on unsupervised-domain-adaptation (UDA) \cite{ganin2016domain,bousmalis2017unsupervised,hoffman2018cycada,mrq15hsj9yang2020fda,tsai2018learning}, which trains on labeled data from the source domain and aims to achieve good performance on data from the target domain without access to the labeled data in the target domain \cite{miller2019simplified}.

Recently, adversarial learning has been widely used in the domain adaptation field \cite{mrq10hsj10zn7zhu2017unpaired,bousmalis2017unsupervised,ganin2016domain,tsai2018learning}.
The primary objective is to instruct a discriminator in differentiating inputs based on their original domain, while the generator concurrently attempts to misdirect it.
% \todo{Chinglish. Rewrite.}
%
Existing methods can be divided into three categories according to the input of the discriminator: image-level alignment \cite{mrq10hsj10zn7zhu2017unpaired,mrq15hsj9yang2020fda,yang2020label,yang2020phase}, feature-level alignment \cite{tsai2018learning,wu2021unsupervised}, and their mixture \cite{hoffman2018cycada}.
Cycle-GAN \cite{mrq10hsj10zn7zhu2017unpaired} breaks the rule of requiring paired cross-domain images in image-to-image tasks and is followed by many subsequent works. 
For example, Yang \emph{et al.} \cite{mrq15hsj9yang2020fda} proposed to minimize the domain discrepancy by exchanging the low-frequency information provided by Fourier transformation of cross-domain images. 
Yang \emph{et al.} \cite{yang2020label} proposed a label-driven module to reduce the image translation bias for improving semantic segmentation performance.
In addition, a number of similar works translate the image style of cross-domain to align the image space \cite{mrq15hsj9yang2020fda,yang2020phase}.

Similar to image-level alignment, feature-level alignment minimizes the domain discrepancy in the feature space.
For instance, Tsai \emph{et al.} \cite{tsai2018learning} constructed a multi-level adversarial network to effectively perform output space domain adaptation at different feature levels.
Wu \emph{et al.} \cite{wu2021unsupervised} constructed a Variational Autoencoder (VAE) to extract modality-invariant latent features.
Following the success of feature-level alignment and image-level alignment, their combination is demonstrated better.
Hoffman \emph{et al.} \cite{hoffman2018cycada} proposed a model to adapt between domains using both generative image space alignment and latent space alignment.
Chen \emph{et al.} \cite{chen2020unsupervised} transformed the appearance of images across domains and enhance domain-invariant of the extracted features for superior performance.

In addition to adversarial learning, UDA in Teacher-Student networks and disentangled representation has also achieved great progress.
Pham \emph{et al.} \cite{pham2021meta} proposed a meta pseudo labels method, which updates the student based on the pseudo-labeled data produced by the teacher and also updates the teacher based on the student’s performance.
As a result, the teacher can generate better pseudo-labels to teach the student.
Different from the disentangled method \cite{pei2021disentangle}, we perform Fourier decomposition to extract the high- and low-frequency portions of an image for domain-invariant and domain-specific information.
These information are employed for segmenting fine-grained domain-invariant structures and synthesizing cross-domain images.

\section{Proposed Method}
\label{S3}
We resolve the domain adaptation task with a novel unsupervised domain adaptation framework, where a cycle-consistent network is proposed to capture the fine-grained structure. Fig. \ref{fig2} shows an overview of the proposed \ourmodel{} for fetal brain tissue segmentation with thick-slice MR scans, which consists of three key components: 1) longitudinal-complete source data construction; 2) Fourier code extraction; and 3) the jointly trained generator and segmentor with cycle consistency.

It should be noted that our model works with thick-slice MR scans, implying that the inter-slice differences are large \cite{zhang2021confidence,huang2021handbook}. Therefore, we design our model as a 2D network, instead of 3D one. 

\subsection{Longitudinal-complete Source Data Construction}
\label{S3-A}

\begin{figure}[t]
  \centering
  \includegraphics[width=\linewidth]{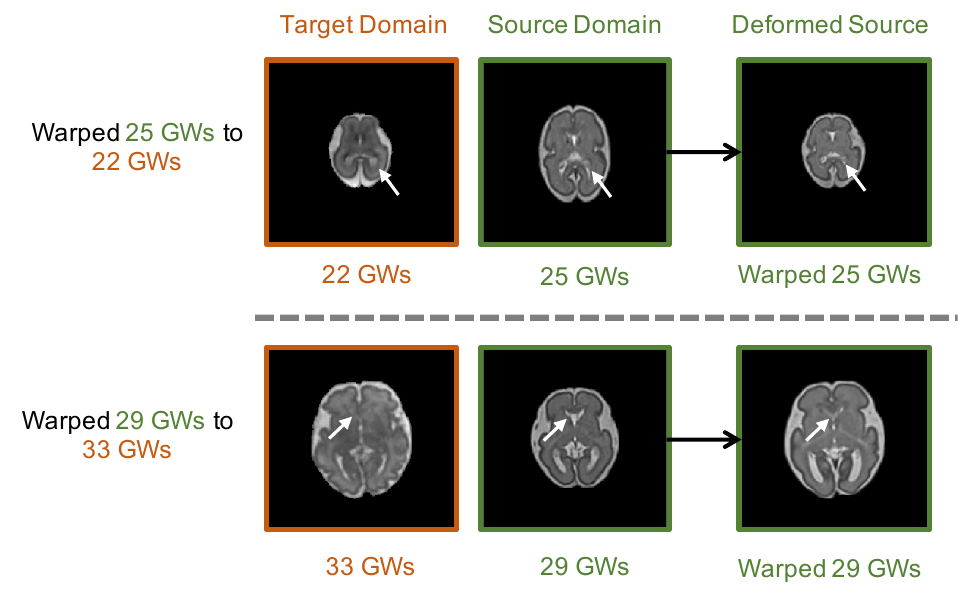}
  \vspace{-20pt}
  \caption{Illustration of the longitudinal-complete source data construction procedure. The source images are registered to the closest GWs target images. The deformed source images and the original source images form longitudinal-complete source data, which were fed to the segmentor.}
  \label{registration}
\end{figure}

To address the issue of severe missing data in longitudinal source data, we register the source images to the target images with close GWs using ANTs toolkit \cite{avants2009advanced} to build deformed longitudinal-complete source data, as shown in Fig. \ref{fig2}. I and Fig. \ref{registration}.
This deformation aligns the source images with the target images, filling in missing GWs and maintaining a similar geometric structure.
We combine the deformed source images with the original source images to create longitudinal-complete source data, enabling the network to learn more about developmental characteristics. Note that the source and target images are not necessarily to be paired for the segmentor, allowing for a flexible and generalizable network that can handle complicated data composition and involve all data in the training process.
\subsection{Fourier Code Extraction}
\label{S3-B}

UDA typically involves training a network to achieve disentanglement of representations.
In this work, we propose to leverage the Fourier transform to efficiently extract disentangled representations, which are referred to as Fourier Style Code (FSC) and Fourier Content Code (FCC).

Specifically, given the MR volume of a fetal brain subject $x\in \mathbb{R}^{H\times W\times N}$, which consists of $N$ slices, we extract its frequency map using the Fast Fourier Transform (FFT) algorithm (i.e., $\mathscr{F}(\cdot)$) \cite{frigo1998fftw}.
Accordingly, $\mathscr{F^{-1}}(\cdot)$ is the Inverse Fast Fourier Transform (IFFT) that maps the frequency signals back to images.
To decompose the frequency signals for extracting style and content codes, we define a binary map (i.e., $M_\alpha$), where the values are all zeros except for the center region determined by a ratio factor $\alpha$:
\begin{equation}
M_\alpha = \begin{cases}
1, & (h,w,n)\in \alpha\times[-H:H, -W:W, -N:N] \\
0, & \text{otherwise}
\end{cases}
\end{equation}
where $\alpha$ is empirically set to 0.05 in our case. Note that the coordinates of the center of $M_\alpha$ is $(0,0)$.
We then apply the $M_\alpha$ to the frequency map to obtain FSC, i.e.,
\begin{equation}
\mathrm{FSC}=\mathscr{F^{-1}}(M_\alpha\circ\mathscr{F}(x)).
\end{equation}
$\circ$ represents the Hadamard product, i.e., element-wise matrix multiplication.
% \todo{should be element-wise multiplication, right?}
Similarly, the FCC is defined as:
\begin{equation}
\mathrm{FCC}=\mathscr{F^{-1}}((\mathbbm{1}-M_\alpha)\circ\mathscr{F}(x)),
\end{equation}
where the $\mathbbm{1}\in \mathbb{R}^{H\times W\times N}$ represents a matrix of all ones.

The procedure of extracting FSC and FCC is illustrated in Fig. \ref{fig2} \uppercase\expandafter{\romannumeral2}.
The FCC captures the content information while discarding domain-specific style information, leading to improved segmentation performance across domains.
Additionally, we incorporate a cycle generation network to better comprehend the fine-grained, domain-invariant anatomy, where FSC provides the domain-specific style information as input to the generator along with the prediction map.
In general, the Fourier code is a reliable and effective method for disentangling domain-invariant content and domain-specific style information, facilitating accurate thick-slice fetal brain tissue segmentation by utilizing high-quality isotropic volumes to their full potential.
The details of the framework are described below.

\subsection{Cycle-consistent Domain Adaptation Network}
\label{Net_section}
\subsubsection{FCC for Domain Adaptive Segmentation}
\label{FCC}
Fig. \ref{fig2}. \uppercase\expandafter{\romannumeral2} shows appearance of the FCC.
As can be observed, FCC can capture the structural details at the boundaries and provides plentiful domain-invariant edge information.

\subsubsection{FSC for Cycle-Consistency}

Cycle consistency learning is a well-known technique and has been widely employed in machine translation \cite{hsj7lis2019detecting}, image synthesis \cite{mrq10hsj10zn7zhu2017unpaired}, etc.

In the field of UDA, Giancarlo \emph{et al.} \cite{hsj7lis2019detecting} proposed a pixel-wise anomaly detection framework to find dissimilarities between input and generated images.
However, the use of only a segmentation map for synthesis can only provide structural information, which may not be sufficient to bridge the domain gap in styles between different domains. In this work, we are inspired by \cite{mrq15hsj9yang2020fda} and consider FSC as a simple style code, which allows our framework to better capture domain-specific style information and achieve improved segmentation performance across domains.

Specifically, for each input 2D MR image \textbf{$x_i$} representing the $i$-th slice in fetal brain subject $x$, our image translation cycle should be able to bring $x_i$ back to the original image. This is achieved through the forward cycle consistency procedure:
$
x_i\to \mathrm{FCC}_i\to S(\mathrm{FCC}_i)\to G(S(\mathrm{FCC}_i), \mathrm{FSC}_i)\approx x_i,
$
where $S(\cdot)$ and $G(\cdot)$ represent the segmentor and generator, respectively. This procedure is illustrated in Fig. \ref{fig2} II and III. The accuracy of the segmentation provides promising image prediction, and vice versa for incorrect segmentation. This consistency principle acts as a constraint for improving the segmentation of fetal brain tissues, and its superior performance will be demonstrated in our ablation study (Section \ref{sec:ablation}).

For the source and target domains, we compute the MSE loss between $G(S(\mathrm{FCC}_i), \mathrm{FSC}_i)$ and $x_i$, denoted as $\mathcal{L}_{\mathrm{syn}}$.
Similarly, for each segmentation result $S(\mathrm{FCC}_i)$, $G$ and $S$ should also satisfy backward cycle consistency:
$
S(\mathrm{FCC}_i)\to G(S(\mathrm{FCC}_i), \mathrm{FSC}_i)\to S(G(S(\mathrm{FCC}_i), \mathrm{FSC}_i)) \approx S(\mathrm{FCC}_i).
$

We further add a discriminator $D(\cdot)$ to encourage the encoder of segmentor $S_\mathrm{E}(\cdot)$ to extract the domain-invariant feature with an adversarial loss defined as:

\begin{align}
    \mathcal{L}_{\mathrm{a d v}}=& \mathbb{E}_{\mathrm{FCC}^{(s)}}\left[\log \left(D\left(S_\mathrm{E}\left(\mathrm{FCC}_{i}^{(s)}\right)\right)\right)\right] \nonumber \\
    &+\mathbb{E}_{\mathrm{FCC}^{(t)}}\left[\log \left(1-D\left(S_\mathrm{E}\left(\mathrm{FCC}_{i}^{(t)}\right)\right)\right)\right].
\end{align}
    
We employ cycle consistency to facilitate self-supervised learning of the target data and fully capture the relationship between the segmentation map and the generated results. This leads to accurate segmentation results at both global and local levels, without relying solely on adversarial learning for UDA.

Using FSC as the style code is a straightforward and effective method for medical images with fixed anatomy and similar content, as demonstrated by our experiments in Section~\ref{sec:ablation}. Also, incorporating FSC improves the performance of both the segmentor and the generator by providing pixel-level constraints.

\begin{figure}[t]
  \centering
  \includegraphics[width=\linewidth]{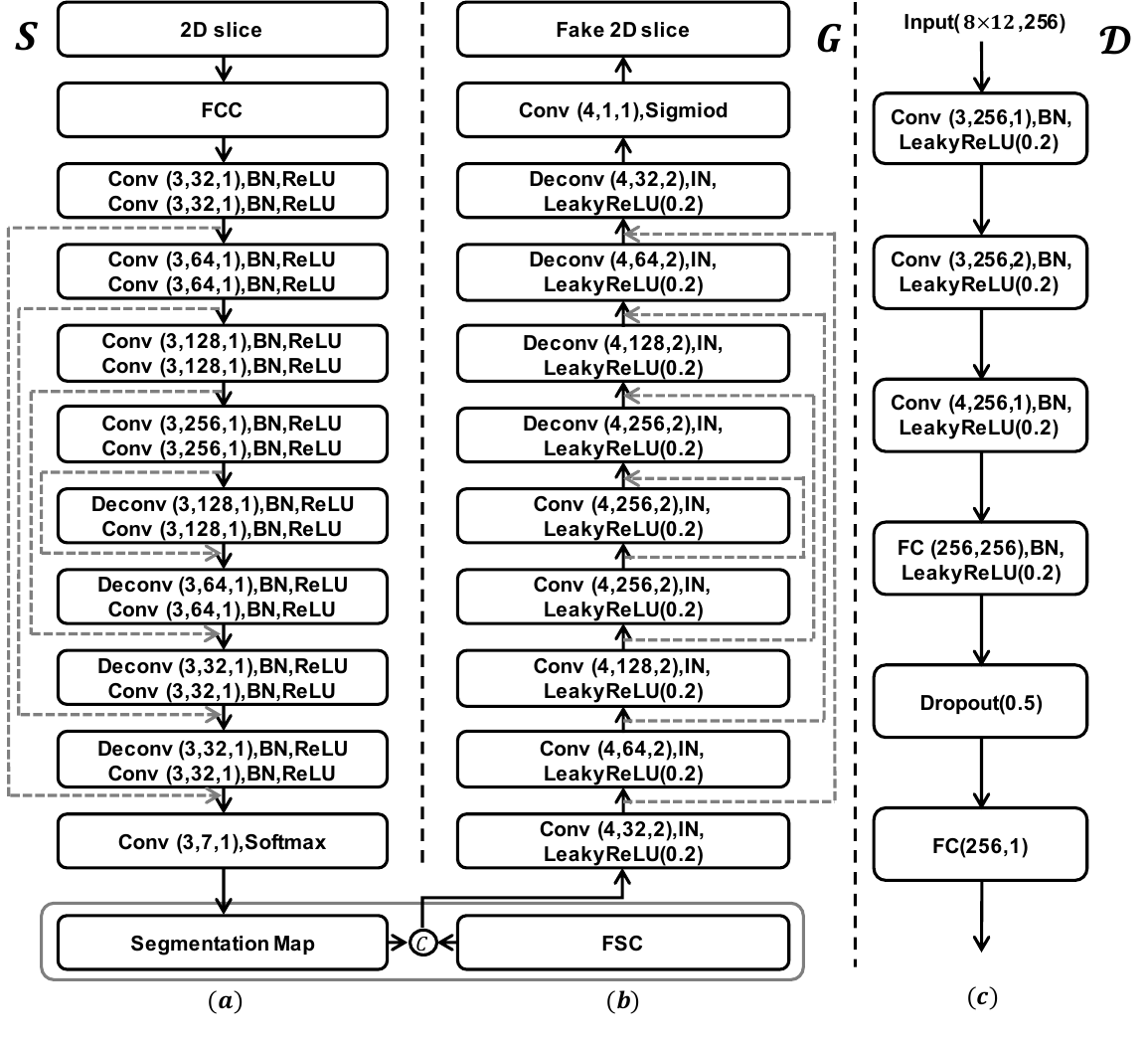}
  \vspace{-20pt}
  \caption{Architectures of segmentor, generator, and the corresponding discriminator. ``$\text{Conv/Deconv}(k,n,s)$'' denotes the convolutional or deconvolutional layer with kernel size ${k}\times {k}$, stride $s$, and $n$ output channels; ``input(a,c)'' represents the size and channels of the input; The sizes of the 2D slice, FCC, segmentation map, and FCC are all $128\times 192$; ``$\text{FC}(n1,n2)$'' represents the fully connected layer with input channel $n1$ and output channel $n2$.}
  \label{fig4}
\end{figure}
\subsubsection{Learning Process}
The segmentation network and generator network are trained in an end-to-end manner. The total loss function $\mathcal{L}$ consists of three components, including the segmentation loss $\mathcal{L}_{\mathrm{seg}}^{(s)}$, the synthetic losses $\mathcal{L}_{\mathrm{syn}}^{(s)}$ (source domain), and $\mathcal{L}_{\mathrm{syn}}^{(t)}$ (target domain) from generator and the adversarial losses $\mathcal{L}_{\mathrm{adv}}$ from $\mathcal{D}(\cdot)$.
Mathematically, $\mathcal{L}$ is defined as follows:
\begin{equation}
\mathcal{L} = \mathcal{L}_{\mathrm{seg}}^{(s)}+\gamma\mathcal{L}_{\mathrm{adv}}+\beta(\mathcal{L}_{\mathrm{syn}}^{(s)}+\mathcal{L}_{\mathrm{syn}}^{(t)}),
\end{equation}
where $\beta=3.0$ and $\gamma=0.1$ are hyper-parameter set empirically.
The generator is discarded at the testing stage, while the FCC images of the target domain directly go through the segmentation network to generate the final segmentation results.

\subsection{Network Configuration and Implementation Details}

\subsubsection{Network Backbone}
The entire network consists of a generator {$G(\cdot)$}, a segmentor {$S(\cdot)$}, and a discriminator {$D(\cdot)$}, each of which is built with a 2D convolutional neural network illustrated in Fig. \ref{fig4}.
Both the segmentor and generator are jointly trained in an end-to-end manner. Note that the input of the segmentor is FCC, which is generated with the FFT implemented using Pytorch.

\subsubsection{Implementation Details}
The proposed method was implemented using the PyTorch platform in Python.
For data preprocessing, we first rotate the original MR image with an arbitrary angle ranging from $0^{\circ}$ to $360^{\circ}$ and an origin at image center. The images are then resized to $128\times 192$ as the input of the segmentation network.
We trained the model on an RTX3060 GPU with a learning rate of $1.0\times 10^{-4}$ using the Adam optimizer to optimize the generator and segmentor parameters. The discriminator is Adam-optimized every three epochs with a learning rate of $1.0\times 10^{-5}$ during the training.

\section{Experiments}
\subsection{Dataset}
% \setParDis
We have collected a large-scale dataset consisting of thick-slice MR scans of 372 prenatal fetuses from maternity hospitals.
The original source data contains 25 reconstructed volumes ranging from 25 GWs to 29 GWs \cite{xia2019fetal} with a voxel size of 0.75$\times$0.75$\times$0.75 (mm).
We transform the original source data using the method described in Section~\ref{S3-A} and update the source data based on registration.
The target data contains 372 thick-slice scans ranging from 22 GWs to 33 GWs with a voxel size of 0.75$\times$0.75$\times$4.4 (mm).
We use all source images, and randomly select 210 target ones for training, 60 target images for validation, and 102 for testing. 
The ground truths of all subjects are annotated by professional doctors.
All images are preprocessed with fetal brain segmentation \cite{zhang2021confidence} and bias field correction \cite{lyx1yushkevich2006user}, and further normalized with z-score.

\begin{table*}[t]
\centering
\setlength\tabcolsep{2pt}
\caption{Quantitative evaluation of the proposed method and competing methods. Ven., Cer. and Bra. are short for Ventricles, Cerebellum and Brainstem, respectively. The best scores are in \textbf{boldface}. $\uparrow$ indicates the higher the score the better and vice versa for $\downarrow$.}
\label{tab1}
\renewcommand{\arraystretch}{1.2}
\begin{tabular}{c|ccccccc|ccccccc} 
\toprule
\multirow{2}{*}{Method} & \multicolumn{7}{c|}{Dice [\%] $\uparrow$}                                                & \multicolumn{7}{c}{ASSD (mm)$\downarrow$}                                              \\ 
\cline{2-15}
                        & CSF      & GM       & WM       & Ven. & Cer. & Bra. & Mean     & CSF     & GM      & WM      & Ven. & Cer. & Bra. & Mean     \\ 
\hline
WoDA                    & 86.5±2.4 & 70.0±4.1 & 88.6±2.6 & 65.9±9.9   & 76.3±21.0  & 57.8±8.5  & 74.2±5.7 & 0.4±0.1 & 0.5±0.2 & 0.5±0.1 & 1.0±0.3    & 3.3±3.0    & 1.7±0.4   & 1.2±0.5  \\
FS                      & 91.9±1.6 & 76.3±3.3 & 91.7±1.1 & 87.6±3.9   & 90.4±4.1   & 84.4±3.7  & 87.0±1.5 & 0.2±0.0 & 0.3±0.0 & 0.3±0.0 & 0.4±0.1    & 0.4±0.1    & 0.6±0.1   & 0.4±0.1  \\ 
\hline
FDA \cite{mrq15hsj9yang2020fda}                & 87.6±2.3 & 68.4±3.5 & 88.9±1.9 & 73.9±6.2   & 71.8±14.3  & 57.4±13.7 & 74.7±5.7 & 0.3±0.0 & 0.4±0.1 & 0.5±0.1 & 0.8±0.2    & 6.5±2.2    & 1.7±0.6   & 1.7±0.4  \\
AdaptSegNet \cite{tsai2018learning}             & 86.8+2.3 & 69.4±3.9 & 88.7±2.0 & 65.1±12.0  & 78.3±21.3  & 59.9±8.1  & 74.7±5.6 & 0.4±0.1 & 0.4±0.1 & 0.5±0.1 & 1.0±0.3    & 2.3±3.1    & 2.1±0.5   & 1.1±0.6  \\ 
CyCADA \cite{hoffman2018cycada}               & 87.6±2.4 & 69.6±4.2 & 89.2±2.1 & 75.3±9.3   & 78.8±20.8  & 59.3±10.3 & 76.6±6.1 & 0.3±0.0 & 0.4±0.1 & 0.5±0.1 & 0.8±0.3    & 4.5±5.8    & 2.0±0.4   & 1.4±1.0  \\
Pham \emph{et al.} \cite{pham2021meta}             & 88.0+2.4 & 72.1±2.9 & 89.2±2.0 & 73.7±9.4  & 80.8±19.7  & 61.3±10.8  & 77.5±5.3 & 0.3±0.0 & 0.4±0.1 & 0.5±0.1 & 0.8±0.3    & 1.1±0.9    & 1.6±0.4   & 0.8±0.2  \\ 
\hline
ANTs \cite{avants2009advanced}             & 76.0+3.8 & 43.8±6.9 & 80.8±3.9 & 52.0±8.7  & 78.9±7.6  & 64.9±8.9  & 66.1±4.4 & 0.6±0.1 & 0.6±0.1 & 1.0±0.2 & 1.3±0.4    & 0.8±0.3    & 1.2±0.5   & 0.9±0.2  \\ 
\hline
\ourmodel{} (Ours)                  & \textbf{89.9±2.2} & \textbf{74.1±2.7} & \textbf{90.5±1.3} & \textbf{83.4±5.5}   & \textbf{88.9±4.8}   & \textbf{78.9±7.1}  & \textbf{84.3±2.1} & \textbf{0.2±0.0} & \textbf{0.3±0.0} & \textbf{0.4±0.1} & \textbf{0.5±0.1}    & \textbf{0.4±0.2}    & \textbf{0.8±0.2}   & \textbf{0.4±0.1}  \\
\bottomrule
\end{tabular}
\end{table*}

\begin{figure*}[t]
  \centering
  \includegraphics[width=\linewidth]{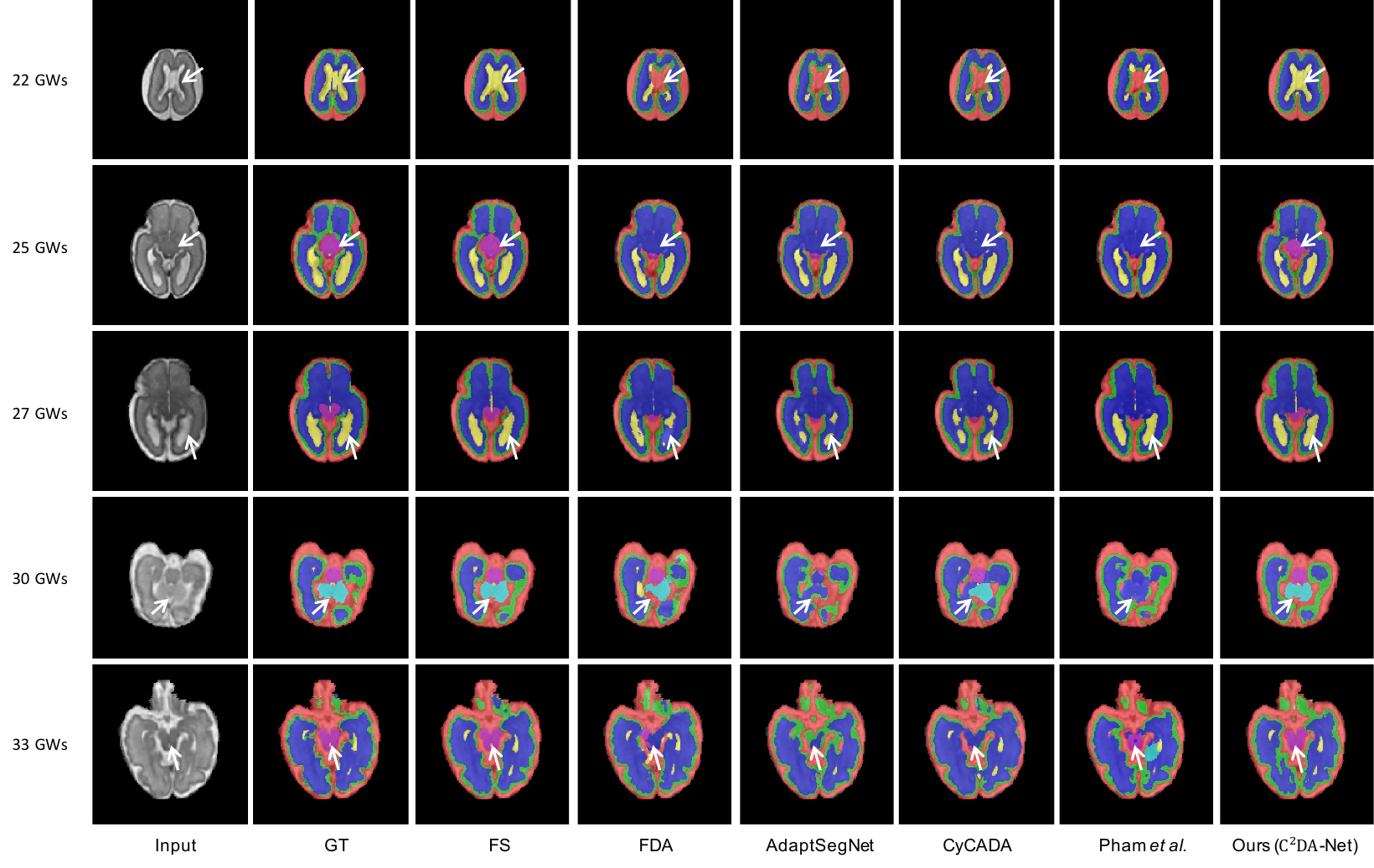}
  \vspace{-20pt}
  \caption{Qualitative comparison of different methods for subjects at 22 GWs, 25 GWs, 27 GWs, 30 GWs, and 33 GWs, respectively. Typical results are shown row-by-row. The red, green, deep blue, yellow, shallow blue and purple stand for CSF, GM, WM, ventricles, cerebellum, and brainstem, respectively.}
  \label{fig6}
\end{figure*}

\begin{figure*}[t]
  \centering
  \includegraphics[width=\linewidth]{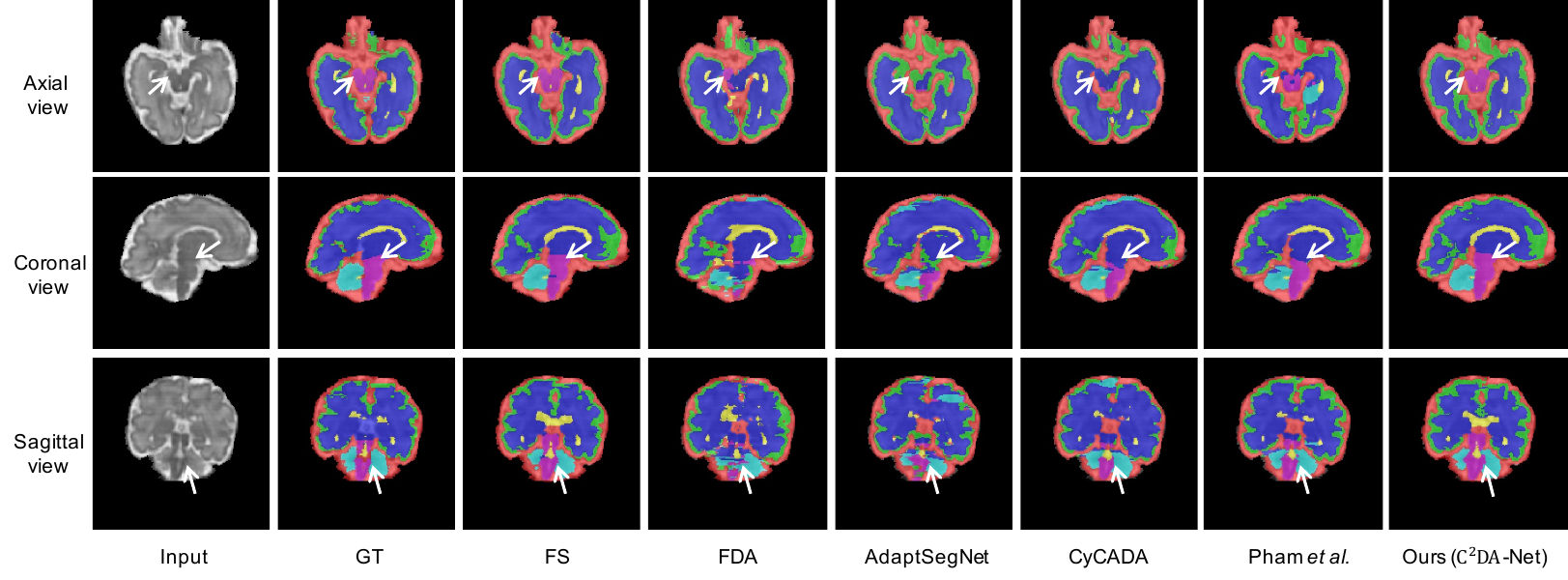}
  \vspace{-15pt}
  \caption{Qualitative results of different methods in three orthogonal views from a single subject. Typical examples are shown row-by-row.}
  \label{fig5}
\end{figure*}

\begin{table*}
\centering
\caption{Quantitative results at different gestational weeks.}
\label{GWs_tab}
\renewcommand{\arraystretch}{1}
\begin{tabular}{c|ccccccc|ccccccc} 
\toprule
\multirow{2}{*}{GWs} & \multicolumn{7}{c|}{Dice {[}\%{]} $\uparrow$}                                & \multicolumn{7}{c}{ASSD (mm)$\downarrow$}                                 \\ \cline{2-15} 
                     & CSF   & GM    & WM    & Ven. & Cer. & Bra. & Mean  & CSF  & GM   & WM   & Ven. & Cer. & Bra. & Mean \\ \hline
22                   & 89.08 & 74.13 & 88.75 & 78.34      & 91.37      & 74.55     & 82.70 & 0.08 & 0.13 & 0.12 & 0.22       & 0.05       & 0.69      & 0.21 \\
23                   & 89.16 & 73.75 & 87.88 & 81.66      & 89.20      & 60.80     & 80.41 & 0.08 & 0.14 & 0.18 & 0.26       & 0.14       & 0.93      & 0.29 \\
24                   & 87.70 & 75.28 & 91.19 & 90.48      & 81.59      & 73.02     & 83.21 & 0.09 & 0.13 & 0.14 & 0.08       & 0.27       & 0.43      & 0.19 \\
25                   & 86.43 & 71.62 & 91.42 & 83.70      & 92.01      & 78.29     & 83.91 & 0.10 & 0.19 & 0.13 & 0.22       & 0.09       & 0.37      & 0.18 \\
26                   & 89.70 & 77.71 & 93.49 & 70.86      & 93.02      & 87.55     & 85.39 & 0.08 & 0.12 & 0.12 & 0.30       & 0.09       & 0.30      & 0.17 \\
27                   & 89.26 & 78.55 & 90.74 & 84.91      & 90.53      & 77.55     & 85.26 & 0.08 & 0.12 & 0.13 & 0.21       & 0.12       & 0.37      & 0.17 \\
28                   & 87.88 & 75.44 & 91.67 & 84.87      & 93.05      & 81.39     & 85.72 & 0.07 & 0.13 & 0.14 & 0.21       & 0.09       & 0.41      & 0.18 \\
29                   & 91.60 & 74.68 & 90.27 & 92.54      & 88.25      & 87.76     & 87.52 & 0.08 & 0.16 & 0.17 & 0.09       & 0.21       & 0.35      & 0.18 \\
30                   & 92.39 & 72.36 & 91.44 & 85.88      & 88.75      & 83.19     & 85.67 & 0.09 & 0.20 & 0.17 & 0.13       & 0.26       & 0.34      & 0.20 \\
31                   & 87.09 & 73.68 & 91.15 & 70.64      & 92.83      & 79.09     & 82.41 & 0.12 & 0.15 & 0.17 & 0.29       & 0.12       & 0.52      & 0.23 \\
32                   & 86.63 & 56.60 & 87.81 & 78.59      & 94.42      & 87.06     & 81.85 & 0.13 & 0.26 & 0.25 & 0.28       & 0.09       & 0.35      & 0.23 \\
33                   & 89.23 & 72.68 & 90.72 & 69.25      & 81.82      & 74.59     & 79.71 & 0.09 & 0.20 & 0.21 & 0.31       & 0.30       & 0.54      & 0.28 \\ \hline
\toprule
\end{tabular}
\end{table*}

\begin{figure}[t]
  \centering
  \includegraphics[width=1.1\linewidth]{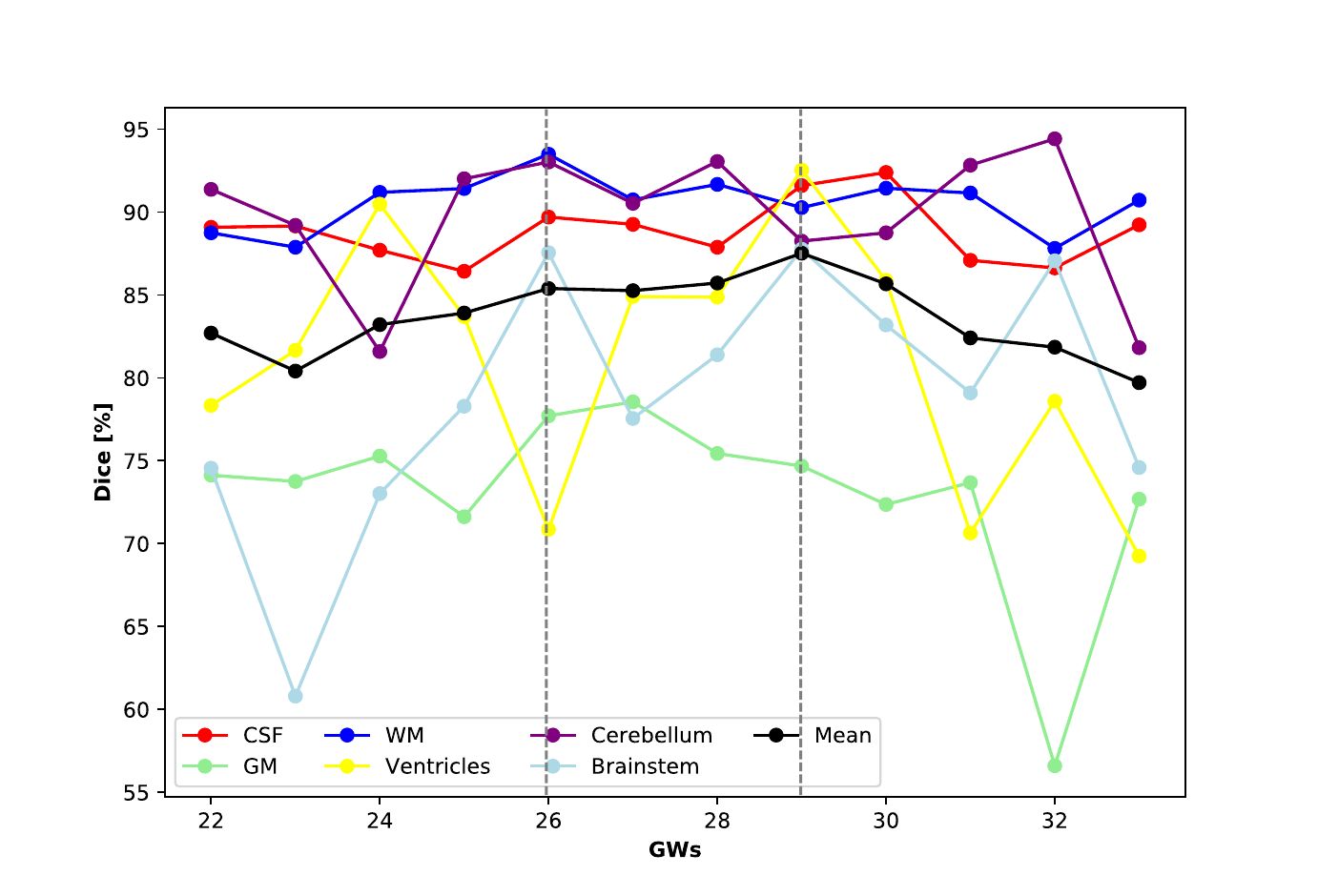}
  \vspace{-20pt}
  \caption{Segmentation performance of our method for six tissues at different GWs, ranging from 22 to 33 GWs, with the original source data ranging from 26 to 29 GWs.}
  \label{GWs}
\end{figure}

\subsection{Experimental Settings and Evaluation Metrics}
\label{4-B}
% \setParDis

We compared our framework with several cutting-edge approaches, including:

\begin{itemize}
    \item \textbf{W/o domain adaptation (WoDA)}: Segmentation is performed with only the basic segmentor, which is considered as the lower bound performance of this task. There is no adversarial learning. The generator module is trained on the source domain and used directly on the target domain without undergoing any adaptation for the target domain.
    \item \textbf{Full supervision (FS)}: This approach is built on the proposed method and is trained on both target and source domains with corresponding ground-truth labels, which is regarded as the upper bound performance.
    \item \textbf{FDA} \cite{mrq15hsj9yang2020fda}, \textbf{AdaptSegNet} \cite{tsai2018learning} and \textbf{CyCADA} \cite{hoffman2018cycada}: These are representative UDA  approaches, where the {FDA} and {AdaptSegNet} align different domains in the image space and feature space, respectively. In contrast, {CyCADA} aligns different domains on both feature and image spaces. 
    \item \textbf{Pham \emph{et al.}} \cite{pham2021meta}: In this approach, the teacher and student networks are updated using pseudo labels to improve the performance of the unannotated data.
    \item \textbf{ANTs} \cite{avants2009advanced}: This approach directly registers the source images to the target images of the nearest GW to get the corresponding label.
    \item \textbf{\ourmodel{} (Ours)}: This is our approach described in Section \ref{S3}.
    
\end{itemize}
For a fair comparison, all the networks utilize the same backbone.

The quantitative comparison is presented in Table \ref{tab1}, where the Dice similarity coefficient (Dice) and the Average Symmetric Surface Distance (ASSD) \cite{dou20173d} are introduced to evaluate the accuracy of the segmentation results. 
We report the mean and standard deviation of the metrics in the format of $mean\pm std$.

\subsection{Comparisons and Analysis}

First, we provide the lower bound (WoDA) and upper bound (FS) performance on the target domain. As shown in Table \ref{tab1}, the main difference between WoDA and FS comes from the intrinsic domain gap between two domains. 
Notably, for tissues with small volumetric sizes such as GM, ventricles, cerebellum, and brainstem, our method significantly outperforms WoDA (e.g., 70.0\% vs.74.1; 65.9\% vs. 83.4\%; 76.3\% vs. 88.9\%; 57.8\% vs. 78.9\%), implying that our network captures fine-grained anatomical structures and is sensitive to small tissues in fetal brain tissue segmentation.
Besides, it can also be observed that, for GM segmentation, the FS only gains 76.3\% Dice accuracy, which is lower than other tissues.
The underlying reason is that the GM has significant variation in different subjects \cite{bethlehem2022brain}, increasing the difficulty of accurate segmentation.

Next, we compare our method with state-of-the-art UDA methods to validate its effectiveness.
The quantitative and qualitative comparisons are presented in Table \ref{tab1}, Fig. \ref{fig6} and Fig. \ref{fig5}.
Our method achieves the best performance in the segmentation of six tissues and multi-view images, demonstrating the advantages of the proposed method in domain adaptation based segmentation.
An interesting observation is that the three state-of-the-art UDA methods provide relatively small performance improvement, which can be explained from two aspects.
First, the limited intensity contrast among different tissues can pose a challenge for methods that align different domains in image space, which can result in limited improvement in low-intensity contrast areas.
Second, the target and source domain data are with the same imaging modality (i.e., T2-weighted MR images), making it difficult to extract domain-invariant features and avoid falling into a local optimum during adaptation.
The low Dice accuracy of ANTs suggests that the registration process is not precise enough for achieving accurate segmentation.
To demonstrate the advantage of our method, we provide five qualitative segmentation samples from different GWs in Fig. \ref{fig6}.
It can be observed that our method matches the ground truth well, especially at the boundaries between brain tissues with limited intensity contrast.

\begin{table*}
\setlength\tabcolsep{1pt}
\centering
\caption{Ablation studies of the key components of our method for fetal brain tissue segmentation.}
\label{tab2}
\setlength\tabcolsep{2pt}
\renewcommand{\arraystretch}{1.2}
\begin{tabular}{c|ccccccc|ccccccc} 
\toprule
\multirow{2}{*}{Method} & \multicolumn{7}{c|}{Dice (\%) $\uparrow$}                                                & \multicolumn{7}{c}{ASSD (mm)$\downarrow$}                                              \\ 
\cline{2-15}
                        & CSF      & GM       & WM       & Ven. & Cer. & Bra. & Mean     & CSF     & GM      & WM      & Ven. & Cer. & Bra. & Mean     \\ 
\hline
WoDA                    & 86.5±2.4 & 70.0±4.1 & 88.6±2.6 & 65.9±9.9   & 76.3±21.0  & 57.8±8.5  & 74.2±5.7 & 0.4±0.1 & 0.5±0.2 & 0.5±0.1 & 1.0±0.3    & 3.3±3.0    & 1.7±0.4   & 1.2±0.5  \\
Reg                     & 89.9±2.3 & 70.2±5.5 & 87.8±4.5 & 75.8±6.8   & 82.3±18.9  & 74.1±8.1  & 80.0±5.0 & 0.3±0.1 & 0.4±0.2 & 0.6±0.2 & 1.0±0.5    & 1.3±1.6    & 1.3±0.7   & 0.8±0.4  \\
Reg-FCC                 & 87.9±2.7 & 72.0±3.8 & 88.4±2.5 & 76.1±5.0   & 85.7±7.7   & 77.0±7.2  & 81.2±2.4 & 0.3±0.0 & 0.4±0.1 & 0.5±0.1 & 0.7±0.1    & 1.0±1.0    & 0.8±0.2   & 0.6±0.2  \\
Reg-G                   & \textbf{90.6±2.2} & 73.4±3.7 & 89.3±2.6 & 82.0±7.3   & 85.6±14.1  & 77.8±7.1  & 83.1±3.3 & 0.2±0.0 & 0.3±0.1 & 0.5±0.1 & 0.5±0.2    & 0.7±0.7    & 0.8±0.2   & 0.5±0.2  \\
Reg-FCC-G               & 88.2±2.6 & 73.0±3.2 & 90.3±1.3 & \textbf{83.7±5.4}   & 87.8±4.1   & 78.6±4.1  & 83.6±2.1 & 0.3±0.0 & 0.3±0.0 & 0.4±0.1 & 0.5±0.2    & 1.7±1.4    & 0.8±0.3   & 0.7±0.3  \\ 
\hline
\ourmodel{}                 & 89.9±2.2 & \textbf{74.1±2.7} & \textbf{90.5±1.3} & 83.4±5.5   & \textbf{88.9±4.8}   & \textbf{78.9±7.1}  & \textbf{84.3±2.1} & \textbf{0.2±0.0} & \textbf{0.3±0.0} & \textbf{0.4±0.1} & \textbf{0.5±0.1}    & \textbf{0.4±0.2}    & \textbf{0.8±0.2}   & \textbf{0.4±0.1}  \\
\bottomrule
\end{tabular}
\end{table*}

\begin{figure*}[t]
  \centering
  \includegraphics[width=\linewidth]{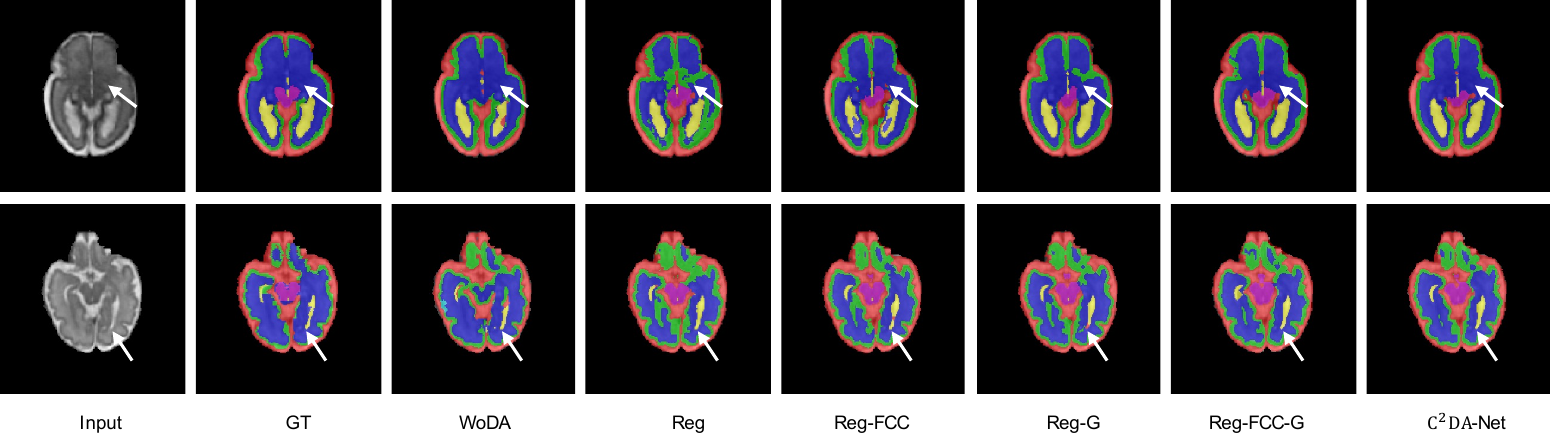}
  % \vspace{-20pt}
  \caption{Qualitative results of the ablation study in two samples. The last five columns correspond to the five controlled experiments in Table \ref{tab2}.}
  \label{ablation}
\end{figure*}

We present further experiments to evaluate the generalization ability of our method. Specifically, we demonstrate that our approach can accurately segment fetal brain tissues even for gestational weeks (GWs) that are not in the source domain. The quantitative and qualitative prediction results for different GWs are presented in Table \ref{GWs_tab} and Fig. \ref{GWs}, respectively. It can be observed that our method achieves high mean Dice accuracy at all GWs, demonstrating the generalization ability of our method. Moreover, the results at GWs within the original source (i.e., 25 GWs to 29 GWs) are consistently better than those at GWs within the deformed source (i.e., 22 GWs to 24 GWs and 30 GWs to 33 GWs), indicating that the variance in GWs is also a crucial domain shift.
This excellent performance indicates that our method \textit{not only} addresses the domain shift between the source and target domains, \textit{but also} effectively adapts to differences in GWs.

\subsection{Ablation Study}\label{sec:ablation}

There are four key components in our method, including 1) registration for source data, 2) FCC extraction, 3) the generator and 3) discriminator for adversarial learning in the segmentor.
As the FSC is a critical component of the generator, and without it, the synthesis would not be valid, we consider the FSC to be a necessary ingredient of the generator.
To validate the effectiveness of the different components proposed in our method, we design the following ablated versions of our method:

\begin{itemize}
    \item \textbf{Reg}: This ablated version only has a segmentor trained on the registered source images which are described in Section \ref{S3-A} and directly applies the trained model to the target data for segmentation.
    \item \textbf{Reg-FCC}: In this ablated version, we augment the Reg with the FCC, so that the segmentor takes the FCC of registered source images as input and applies the trained model to the FCC of target data for segmentation.
    \item \textbf{Reg-G}: This ablated version augments the Reg with a generator, and the segmentor takes the registered source images and target images as input.
    The generator accepts the combination of prediction and the corresponding FSC as input to synthesize the original image. Note that the generator is discarded during the testing stage.
    \item \textbf{Reg-FCC-G}: We augment the Reg-G with the FCC in this ablated version. The segmentor takes the FCC of registered source and target data as input.
    \item \textbf{\ourmodel{} (Reg-FCC-G-Adv)}: This is the full version of our method, which combines all the four key components together.
\end{itemize}

We also include the low bound (i.e., WoDA in Section \ref{4-B}) for better evaluation of different ablated versions. 

\subsubsection{Effectiveness of Registration}
The source data building based on registration is the key technique to tackling the challenge of the domain gap between GWs and is the reason why the ablated version ``Reg'' achieves promising improvement, as shown in Table \ref{tab2}.
In our original source data, there are several missing GWs compared with the target data, which is a common issue in longitudinal fetal MRI.
However, the development characteristics of the fetal brain are mainly reflected in the ventricles, cerebellum, and brainstem at these GWs, thus WoDA method works worse on these three tissues.
After registration, the domain gap caused by the missing GWs is reduced and the segmentation performance of the aforementioned tissues has greatly improved.
Compared to ``WoDA'', ``Reg'' boosts the Dice accuracy of ventricles, cerebellum and brainstem by a large margin of 9.9\%, 6.0\% and 16.3\%, respectively. 

\subsubsection{Effectiveness of FCC}
As described in Section \ref{FCC}, the domain-invariant anatomical structure provided by FCC is a key element of our method and can improve the robustness of segmentation performance.
As listed in Table \ref{tab2}, by taking FCC as input of the segmentor, the standard deviation of both Dice and ASSD results has reduced significantly,
especially for cerebellum, where ``FCC'' decreases the Dice standard deviation of ``Reg'' and ``Reg-G'' by 11.2\% and 10.0\%, respectively. 
Experiments have shown that the FCC effectively captures subtle frequency variations at the boundaries and supplies a substantial amount of domain-invariant information in the image space. 
This advantage benefits the segmentor in learning domain-invariant features and enhances its stability and generalizability in the segmentation of fetal brain tissue.

\subsubsection{Effectiveness of Generator}
The generator is a crucial component for correcting the fine-grained segmentation error, especially in regions where there is no clear intensity difference between adjacent tissues. The use of the generator in our method improves the Dice accuracy of both ``Reg-G'' and ``Reg-FCC-G'' by 3.1\% and 2.4\%, respectively.
In particular, the ventricles benefit greatly from the addition of the generator, with improvements of 6.2\% and 7.6\%, respectively.
A notable observation is that the generator improves the Dice accuracy for every tissue, indicating its effectiveness in capturing anatomically-correct and domain-invariant brain structures.

\subsubsection{Effectiveness of Adversarial Training}

Adversarial learning is a powerful technique that we use in \ourmodel{} to align the features extracted from source and target domain data. This technique has been demonstrated to be effective in various studies \cite{hoffman2018cycada,mrq10hsj10zn7zhu2017unpaired,bousmalis2017unsupervised,ganin2016domain,tsai2018learning} and our results in Table~\ref{tab2} and Fig.~\ref{ablation} confirm its effectiveness in our method.

\subsection{Component of the Fourier Code}

\begin{figure}[t]
  \centering
  \includegraphics[width=\linewidth]{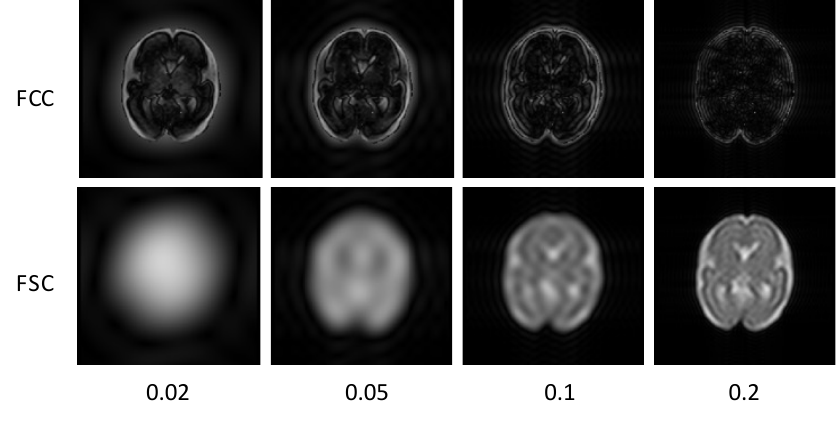}
  \vspace{-20pt}
  \caption{Visualization of FCC and FSC under different $\alpha$ values.}
  \label{fig7}
\end{figure}

In the Fourier code described in Section \ref{S3-B}, the hyperparameter $\alpha$ plays a crucial role in determining the information of FCC and FSC.
To investigate the impact of $\alpha$, we set $\alpha$ to 0.02, 0.05, 0.1, and 0.2, respectively. 
As illustrated in Fig.~\ref{fig7}, a smaller $\alpha$ provides a clearer FCC and a more blurry FSC, and vice versa. 
However, since the generator relies on the FSC to generate realistic synthetic images, it is crucial that the FCC and FSC reach a balance between texture/structure and domain-invariant information. Therefore, choosing an appropriate $\alpha$ is essential for achieving optimal performance. In our experiments, we found that the best results are obtained with $\alpha=0.05$, which thus serves as a default setting in our method.

\section{Discussion}
In this work, we propose a novel registration framework to address the challenge of multi-GW image segmentation in fetal brain MRI.
We introduce the use of Fourier code to represent the domain-invariant structure and the domain-specific style, which is then fed into a cycle-consistency network for anatomically-correct tissue segmentation.
We provide a comprehensive analysis of the proposed method, highlighting the benefits of utilizing isotropic reconstructed high-quality target domain data, the effectiveness of GWs adaptation in fetal brain tissue segmentation, and the significance of Fourier code in segmentation tasks.

\subsubsection{Exploit the Feta 2021 Dataset as the Target domain}
\begin{table}[t]
\centering
\setlength\tabcolsep{1pt}
\caption{Segmentation performance of our method under different $\alpha$.}
\label{tab3}
\renewcommand{\arraystretch}{1.2}
\begin{tabular}{c|ccccccc} 
\toprule
\multirow{2}{*}{$\alpha$} & \multicolumn{7}{c}{Dice [\%] $\uparrow$}                                                  \\ 
\cline{2-8}
                        & CSF      & GM       & WM       & Ven. & Cer. & Bra. & Mean      \\ 
\hline
0.02                    & 86.7±2.8 & 72.6±4.5 & 88.7±3.4 & 82.1±6.6   & 84.5±11.7  & 76.8±7.4  & 81.9±3.1  \\
0.05                    & 89.9±2.2 & \textbf{74.1±2.7} & \textbf{90.5±1.3} & 83.4±5.5   & \textbf{88.9±4.8}   & \textbf{78.9±7.1}  & \textbf{84.3±2.1}  \\
0.1                     & \textbf{90.5±2.0} & 73.4±3.4 & 89.8±1.7 & \textbf{83.5±4.8}   & 87.7±5.2   & 78.4±7.4  & 83.9±1.8  \\
0.2                     & 88.2±1.6 & 68.5±4.0 & 88.0±3.2 & 80.5±3.0   & 85.8±4.7   & 78.9±4.0  & 81.7±1.8  \\ 
\bottomrule
\end{tabular}
\end{table}
\begin{table}[t]
\centering
\setlength\tabcolsep{1pt}
\caption{Quantitative experimental results for the target domain data in Feta 2021.}
\setlength\tabcolsep{1pt}
\renewcommand{\arraystretch}{1.2}
\label{discuss}
\begin{tabular}{c|ccccccc} 
\toprule
\multirow{2}{*}{Method}     & \multicolumn{7}{c}{Dice [\%] $\uparrow$}                                                                                                                                 \\ 
\cline{2-8}
                            & CSF                  & GM                   & WM                   & Ven. & Cer. & Bra.            & Mean                  \\ 
\hline
2D (Ours)                      & 90.3±1.5             & 72.9±5.4             & 89.5±2.0             & 81.0±5.9             & 89.6±4.8             & 78.1±6.5             & 83.6±1.6              \\
2D (FS)                       & 92.9±1.1             & 77.9±4.0             & 92.8±1.3             & 87.6±3.6             & 93.1±2.1             & 90.8±1.8             & 89.2±1.0              \\ 
\hline
3D (Ours)   & 88.0±2.3  & 73.8±3.5  & 90.2±1.3  &84.9±5.1  & 89.6±3.1  & 80.7±4.6  & 84.5±0.9   \\
3D (FS) & 95.2±0.6  & 82.5±4.1  & 94.8±1.2  & 91.8±2.8  & 94.7±1.8  & 94.2±2.0  & 92.2±1.4   \\

\bottomrule
\end{tabular}
\end{table}

To investigate the generalization ability of our method, we perform additional experiments with the data from Feta 2021 challenge dataset, which consists of 80 T2 fetal brain scans reconstructed using two different methods \cite{tourbier2015efficient,kuklisova2012reconstruction}.

Since both the source and target domains are isotropically reconstructed high-quality volumes in this dataset, we extend our framework to the 3D version for optimal performance.
We train our \ourmodel{} from scratch using 25 MR volumes from source domain (25-29 GWs) and 50 MR volumes from target domain (21-35 GWs).
We compare the 2D and 3D versions of \ourmodel{} (i.e., 2D-Net and 3D-Net) by testing on 15 target MR volumes ranging from 21 to 35 GWs.
In addition, we train our 2D and 3D methods on labeled data from both source and target domains, which provides the upper bound performance.
As shown in Table \ref{discuss}, the mean Dice accuracy of 2D-Net is 83.6\%, which is comparable with 3D-Net (84.5\%).
This implies that the 3D network tends to exacerbate the domain shift induced by development differences and image styles at the 3D level.
On the other hand, 2D networks are more efficient and less sensitive to overfitting.
Hence, our method based on 2D network is robust for the domain adaptation task, especially for clinically-acquired thick-slice scans.
% \todo{What is the evidence to support this conclusion?}

\subsubsection{GW Adaptation and Fourier Code}
Segmenting fetal brain tissue is a challenging task due to large variation of anatomical change across different GWs.
By utilizing deformed source images, our method outperforms state-of-the-art techniques, which rely on style or feature adaptation.
Specifically, direct registration can perform well in tissue matching, as shown in Table \ref{tab1} (i.e., ANTs), and the deformed source data can improve performance significantly as shown in Table \ref{tab2} (i.e., Reg).
This demonstrates that the deformed source data can improve performance without strict alignment with the target data.

In Section \ref{FCC}, we argue that the FCC represents domain-invariant structure as it lacks the style feature. In contrast, the FSC contains the style feature and enables the generator to synthesize images of different domains. Based on comparison results in Table \ref{tab2}, we can draw the following conclusions:
1) The FCC is effective in improving the robustness of segmentation, demonstrating the domain-invariant characteristics of FCC, and 2) the FSC can efficiently incorporate style code into the generator's input to ensure cross-domain cycle consistency, making it a practical and effective approach.

\section{Conclusion}
In this paper, we have proposed a novel cycle-consistent domain adaptation network, \ourmodel{}, which can leverage a small set of annotated isotropic volumes to guide tissue segmentation of clinically-acquired thick-slice scans.
In \ourmodel{}, we first construct the longitudinally-complete source data to adapt to the target data distributed across various GWs.
Then, Fourier transformation is introduced to extract domain-invariant (i.e., FCC) and domain-specific (i.e., FSC) information, which are used for effective training.
Finally, the integration of a generator enables our network to operate in a self-supervised manner, making the segmentor capture anatomically-correct cycle consistency at both image and feature levels.
Experimental results on a clinical dataset indicate that our \ourmodel{} is effective in fetal brain tissue segmentation and outperforms the state-of-the-art methods.

\section{Acknowledgement}
This work was supported in part by National Natural Science Foundation of China (grant number 62250710165), and Science and Technology Commission of Shanghai Municipality (STCSM) (grant number 21010502600).
% \clearpage
% \newpage

% \usepackage[pagebackref,breaklinks,colorlinks]{hyperref}

\bibliography{reference}
\bibliographystyle{IEEEtran}

\end{document}